\documentclass[pre,amsmath,amssymb,preprint,superscriptaddress]{revtex4}

\usepackage{amsmath}    
\usepackage{amsfonts}
\usepackage{amssymb}
\usepackage{graphicx}   
\usepackage{dcolumn}
\usepackage{bm}

\begin{document}

\title{Hierarchy of instabilities for two counter-streaming magnetized pair beams}

\author{A. Bret}
\affiliation{ETSI Industriales, Universidad de Castilla-La Mancha, 13071 Ciudad Real, Spain}
 \affiliation{Instituto de Investigaciones Energ\'{e}ticas y Aplicaciones Industriales, Campus Universitario de Ciudad Real,  13071 Ciudad Real, Spain.}

\date{\today }

\begin{abstract}
The instabilities triggered when two counter-streaming pair beams collide are analyzed. A guiding magnetic field is accounting for, while both beams are considered identical and cold. The instability analysis is conducted over the full \textbf{k}-spectrum, allowing to derive the hierarchy map of the dominant unstable modes, in terms of the initial beams energy $\gamma_0$ and a magnetic field strength parameter $\Omega_B$. Four different regions of the $(\Omega_B,\gamma_0)$ phase space are identified, each one governed by a different kind of mode. The analysis also unravels the existence of a ``triple point'', where 3 different modes grow exactly the same rate. A number of analytical expressions can be derived, either for the modes growth-rates, or for the frontiers between the 4 regions.
\end{abstract}

\maketitle

\section{Introduction}
Counter-streaming instabilities are one of the oldest topic in plasma physics \cite{Mikhailovskii1}. When two collisionless plasmas pass through each other, many instabilities are triggered as the unstable spectrum in the \textbf{k}-space is far from monochromatic \cite{BretPoPReview}. At non-relativistic velocities, the two-stream instability tends to be the fastest growing one, which is why the system exhibits it. In the relativistic regime, the unstable spectrum becomes richer, and many different instabilities are likely to govern the linear phase of the system evolution \cite{BretPRL2008}.

Therefore, the knowledge of the outcome of the linear phase comes down to the knowledge of the hierarchy of unstable modes. Given the initial parameters of the system, which mode will grow faster? This problem has already been solved for the case of two counter-streaming electron beams, exhibiting a non-trivial competition between the two-stream, the Weibel and the oblique instabilities \cite{BretPRL2008}. The same system has been analyzed accounting for the ion motion, displaying an even more intricate hierarchy map \cite{bretApJ2009}.

As long as the system in un-magnetized, a pair beam, for example, is equivalent to an electron beam because the equations governing the evolution only incorporate the square of the charges. As a result, the aforementioned studies related to unmagnetized counter-streaming electron beams can be straightforwardly extended to counter-streaming pair beams. In turn, the inclusion of an external magnetic field introduces cyclotron frequencies which depend on the sign of the charges. The consequence is that former works on magnetized counter streaming electron beams \cite{Godfrey1975,BretPoPMagne} cannot be generalized to magnetized pair beams.

In view of the importance of counter-streaming pair beams in astrophysics or collisionless shocks physics for example \cite{SilvaApJ,Spitkovsky2005,Stockem2006ApJ,Keshet2009,SironiApj,DieckmannPPCF2009,Kino2012,BretPoP2013,BretPoP2014,sarri2015}, it would be desirable to know more about the behavior of such systems in the magnetized case.

This is why this article is devoted to elucidating the hierarchy of unstable modes, for two counter-streaming pair beams with a flow-aligned magnetic field. As a first step towards the resolution of the full problem, we consider cold, symmetric beams with identical initial Lorentz factor $\gamma_0$. The amplitude of the magnetic field and $\gamma_0$ are therefore the sole free parameters.

The four-fluids formalism \cite{BretPoPFluide} implemented is described in Section \ref{sec:sec1}, together with the derivation of the dispersion equation. Then in Section \ref{sec2}, the hierarch map of the dominant mode in terms of ($\Omega_B,\gamma_0$) is directly introduced numerically ($\Omega_B$ measures the strength of the field, see Eq. (\ref{eq:var}) below). Given the number of unstable modes involved (5, at least), this strategy allows to focus directly on the dominant ones. The hierarchy map unravels 4 different regions of the ($\Omega_B,\gamma_0$) phase space, where 4 different modes dominate the unstable spectrum. Each mode is analyzed in Section \ref{sec3}. The frontiers between these regions are studied in Section \ref{sec4}, and various analytical expressions are derived before we reach our conclusion.

\section{Formalism}\label{sec:sec1}
We consider therefore two symmetric counter-streaming pair beams with initial velocity $\pm v_0 \mathbf{e}_z$, and common Lorentz factor $\gamma_0=(1-v_0^2/c^2)^{-1/2}$. Both beams are initially cold and embedded in a static magnetic field $B_0 \mathbf{e}_z$. In each beam, the electronic density is $n_0$, identical to the positronic density. We also define the plasma frequency $\omega_p^2=4\pi n_0 q^2/m$, where $q$ and $m$ are the particles' charge and mass. We thus have 4 species involved in the system: the electrons and the positrons from the rightward beam, and the electrons and the positrons from the leftward beam.

The dispersion equation is obtained assuming harmonic perturbations of all quantities of the form $\exp (\imath \mathbf{k}\cdot \mathbf{r} - \imath\omega t)$, before linearizing the 4 matter conservation and the 4 momentum conservation equations. Since the system is symmetric around the $z$ axis, we can chose $\mathbf{k} = (k_x,0,k_z)$ without loss of generality. Such a general choice for the wave-vector is mandatory if one is to capture successfully the most unstable mode.

The matter conservation equations read,
\begin{equation}\label{eq:matter}
  \frac{\partial n_i}{\partial t} + \nabla \cdot (n_i \mathbf{v}_i) =0,
\end{equation}
where $n_i,\mathbf{v}_i$, $i=1\ldots 4$, stand for the densities and velocities of the 4 species involved.  The momentum conservation equations read,
\begin{equation}\label{eq:momentum}
  \frac{\partial \mathbf{p}_i}{\partial t} + (\mathbf{v}_i\cdot\nabla) \mathbf{p}_i = q_i \left( \mathbf{E} + \frac{\mathbf{v}_i \times (\mathbf{B} + \mathbf{B}_0)}{c} \right).
\end{equation}
Once linearized, Eqs. (\ref{eq:matter}) allow to express the first order density perturbations $n_{1,i}$ in terms of the first order velocity perturbations $\mathbf{v}_{1,i}$. These expressions are then in turn introduced in the linearized Eqs. (\ref{eq:momentum}), to obtain the first order velocity perturbations in terms of the first order electric field $\mathbf{E}_1$. Doing so, one needs to use $\mathbf{B} = (c/\omega)\mathbf{k}\times \mathbf{E}$ in order to eliminate the self-generated magnetic field.

Knowing the 4 perturbed velocities in terms of $\mathbf{E}_1$ allows to express the first order current $\mathbf{J}_1$ as,
\begin{equation}\label{eq:J1}
  \mathbf{J}_1 = \sum_{i=1}^4 q_i n_0 \mathbf{v}_{1,i} + \sum_{i=1}^4 q_i n_{1,i} \mathbf{v}_{0,i} \equiv \mathbf{J}_1 (\mathbf{E}_1).
\end{equation}
Maxwell-Faraday’s and Maxwell-Amp\`{e}re’s equations are then combined, yielding
\begin{equation}\label{eq:Max}
 \mathbf{k} \times (\mathbf{k} \times \mathbf{E}_1) + \frac{\omega^2}{c^2}\left(\mathbf{E}_1 + \frac{4 \imath \pi}{\omega} \mathbf{J}_1 \right) \equiv \mathcal{T} (\mathbf{E}_1) = 0.
\end{equation}
The dispersion equation eventually reads $\det \mathcal{T}(\mathbf{k},\omega)=0 $, with the tensor $\mathcal{T}$ defined above. The analytical evaluation of this tensor has been performed using the \emph{Mathematica} Notebook described in \cite{BretCPC}. The tensor elements are reported in Appendix \ref{app} in terms of the dimensionless variables,
\begin{equation}\label{eq:var}
x = \frac{\omega}{\omega_p},~~\mathbf{Z} = \frac{\mathbf{k} v_0}{\omega_p},~~\beta = \frac{v_0}{c},
~~\Omega_B = \frac{\omega_b}{\omega_p},~~\mathrm{with}~~\omega_b = \frac{|q| B_0}{mc}.
\end{equation}

\section{Hierarchy overview}\label{sec2}
The number of unstable modes under scrutiny is quite important, as we scan the full \textbf{k}-spectrum under the variations of our 2 free parameters $\Omega_B$ and $\gamma_0$.

Instead of studying each and every unstable mode, we therefore start by computing the map of the dominant modes in the $(\Omega_B,\gamma_0)$ space. We will then focus only on those modes which get to dominate part of the map.

\begin{figure}
  \includegraphics[width=\textwidth]{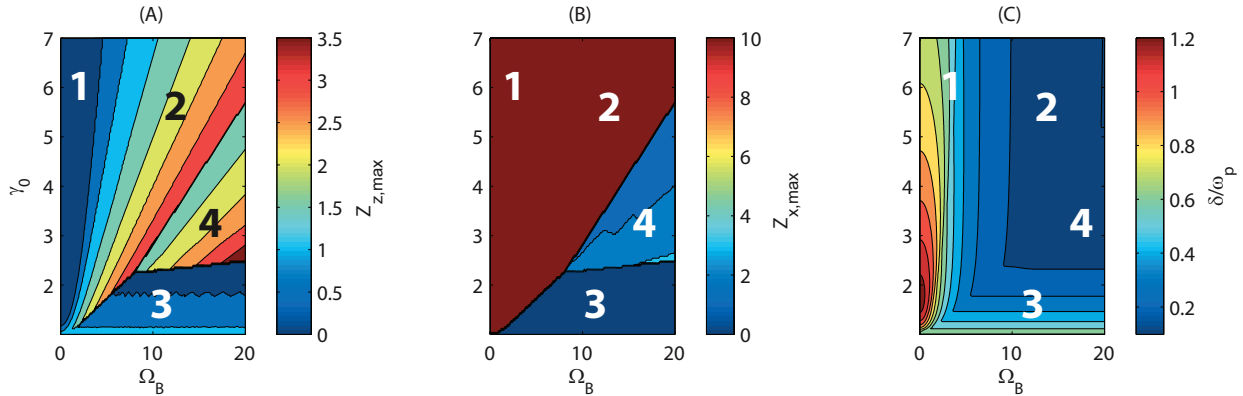}
  \caption{(Color online) Hierarchy map in the $(\Omega_B,\gamma_0)$ space. (A) Parallel component of the most unstable mode. (B) Perpendicular component of the most unstable mode. (C) Growth rate of the most unstable mode. Four regions, numbered 1 to 4, clearly stand out.}\label{fig:hie}
\end{figure}

\begin{figure}
  \includegraphics[width=\textwidth]{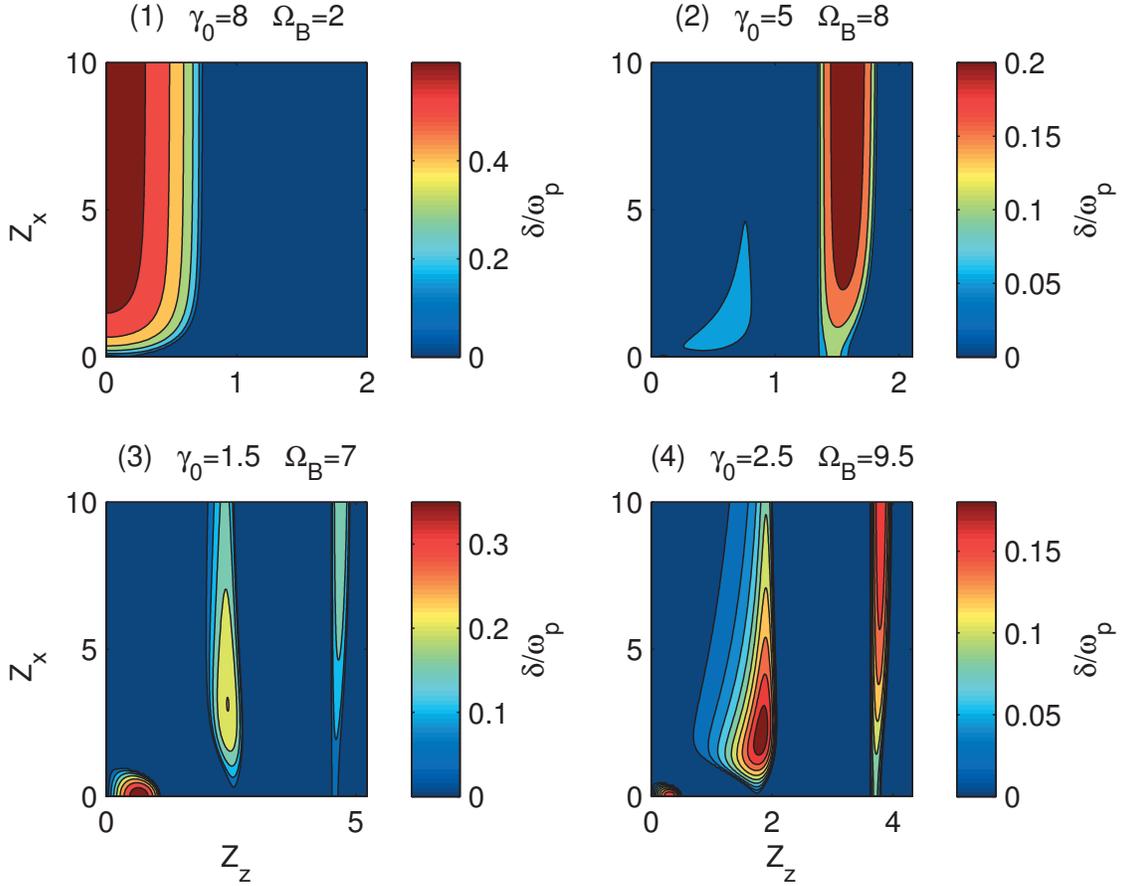}
  \caption{(Color online) Growth rate maps $\delta(\mathbf{Z})$ representative of the spectrum in each of the 4 regions determined by Figs \ref{fig:hie}.}\label{fig:maps}
\end{figure}

This hierarchy map is displayed on Figures \ref{fig:hie}(A-C). The map can clearly be split in 4 regions, and 4 growth-rate maps $\delta(\mathbf{Z})$ typical of each region, are displayed on Figures \ref{fig:maps}(1-4).

Frontiers are easier to detect looking at the components of the fastest growing mode, because transitions here may be discontinuous. The frontier 1/2, between Regions 1 and 2, is best evidenced when looking at the $Z_{\parallel, max}=Z_{z, max}$ map. In turn, frontiers 2/4 and 3/4 clearly stand out on the $Z_{\perp, max}=Z_{x, max}$ map.

In regions 1 and 2, the reader will observe that the fastest growing mode has $Z_x=Z_\perp = 10$. Such modes simply have their growth-rate saturating at large $Z_x$. Since we searched up to $Z_x = 10$, the algorithm returns ``10'' as the normal component of the fastest growing mode. A rigourous answer would be $Z_x = +\infty$. In practice, saturation is reached for $Z_x \gtrsim 5$ (see for example Figs. \ref{fig:maps}-1 and 2). The 4 regions are,

\begin{description}
  \item[Region 1] The fastest growing mode in this region has $Z_z=Z_\parallel =0$ and $Z_x=Z_\perp = 10$. As evidenced on Fig. \ref{fig:maps}-1, the Weibel instability here governs the linear phase.
  \item[Region 2] The fastest growing mode in this region has $Z_z \neq 0$ and $Z_x =10$. As evidenced on Fig. \ref{fig:maps}-2, modes which were dubbed ``upper-hybrid-like'' in Ref. \cite{Godfrey1975}, here govern the linear phase.
  \item[Region 3] The fastest growing mode in this region has $Z_z \neq 0$ and $Z_x = 0$. The analysis conducted later in Section \ref{sec:OTS} shows that there are 2 unstable modes for such wave-vectors. But the dominant instability here is the two-stream instability.
  \item[Region 4] The fastest growing mode in this region has both $Z_z \neq 0$ and $Z_x \neq 0$. As evidenced on Fig. \ref{fig:maps}-4, oblique modes  here govern the linear phase.
\end{description}

We now turn to a region-dependant study of each dominant mode.

\section{Dominant modes in each region}\label{sec3}

\subsection{Region 1: Weibel Region}
With $Z_\parallel=0$, the tensor $\mathcal{T}$ turns diagonal and the tensor elements (\ref{eq:Telt}) read,
\begin{eqnarray}
  \mathcal{T}_{11}&=& 1-\frac{4 \gamma_0}{\gamma_0^2 x^2-\Omega_B^2},  \nonumber \\
  \mathcal{T}_{22}&=& 1-\frac{4 \gamma_0}{\gamma_0^2 x^2-\Omega_B^2}-\frac{Z_x^2}{\beta ^2 x^2}, \nonumber \\
  \mathcal{T}_{33}&=& 1-\frac{1}{x^2} \left[ \frac{4}{\gamma_0^3}+\frac{Z_x^2}{\beta ^2}+\frac{4 \gamma_0 Z_x^2}{\gamma_0^2 x^2 - \Omega_B^2} \right]. \nonumber
\end{eqnarray}
The equation $\mathcal{T}_{11}=0$ clearly cannot yield any solution with $x^2<0$. Regarding $\mathcal{T}_{22}=0$, the equation can also be solved exactly, and a little algebra can prove that its solutions have $x^2>0$ provided $(\beta \gamma_0 \Omega_B Z_x)^2>0$, which is always true. The unstable mode arises therefore from $\mathcal{T}_{33}=0$. This equation is amenable to the polynomial equation,
\begin{equation}\label{eq:PolyWeibel}
\beta^2 \gamma_0^5 x^4-\gamma_0^2 x^2 \left[\beta^2 \left(\gamma_0 \Omega_B^2+4\right)+\gamma_0^3 Z_x^2\right]+\gamma_0^3 Z_x^2 \left(\Omega_B^2-4 \beta^2 \gamma_0\right) + 4 \beta^2 \Omega_B^2= 0,
\end{equation}
which can be solved exactly. Also, the large $Z_x$ value of the growth-rate can be straightforwardly derived by noting that the dispersion function for $Z_x=\infty$ is nothing but the coefficient of $Z_x^2$ in the equation above. We thus find for the maximum Weibel growth-rate $\delta_W$ a result already derived in Ref. \cite{Stockem2006ApJ},
\begin{equation}\label{Eq:Weibel}
  \delta_W=\sqrt{\frac{4 \beta ^2 \gamma_0-\Omega_B^2}{\gamma_0^2}},
\end{equation}
which vanishes for \cite{Stockem2006ApJ,StockemPPCF},
\begin{equation}\label{Eq:Weibel0}
  \Omega_B > 2\beta \sqrt{\gamma_0},
\end{equation}
similar to the threshold found for counter-streaming electron beams \cite{Godfrey1975,BretPoPMagne}.

\begin{figure}
  \includegraphics[width=\textwidth]{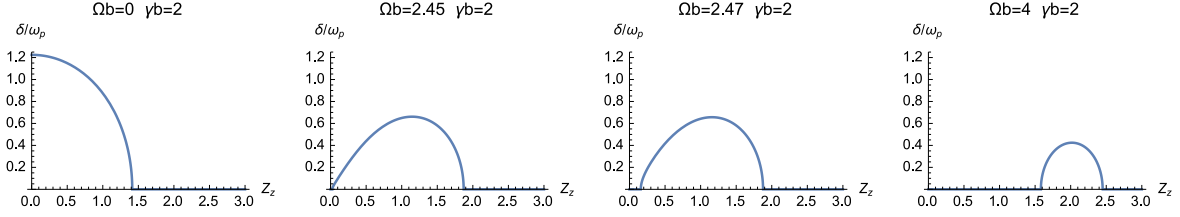}
  \caption{Transition from the Weibel instability to the upper-hybrid-like instability at $Z_x=\infty$, with increasing magnetic field parameter and $\gamma_0$. In accordance with Eq. (\ref{Eq:Weibel0}), the Weibel instability vanishes for $\Omega_B=2\beta\sqrt{\gamma_0}=2.45$.}\label{fig:weibeluhl}
\end{figure}

\subsection{Region 2: Upper-hybrid-like Region}
When the magnetic field is switched on, the Weibel instability progressively decreases \cite{Stockem2006ApJ}, while the most unstable mode at large $Z_x$ migrates towards $Z_z \neq 0$ as displayed on Figures \ref{fig:weibeluhl}. As can be seen on Fig. \ref{fig:maps}-2, upper-hybrid-like modes also have their growth-rate saturating at $Z_x=Z_\perp=\infty$. The dispersion equation for $Z_x$ infinite can be derived from the method described in the preceding subsection. Writing the explicit dispersion equation under the form of a fraction, one starts expressing it in terms of a polynomial equation, where the polynomial is the numerator of the fraction. Then, the dispersion function for $Z_x=\infty$ is the coefficient of $Z_x^n$ in this polynomial, where $n$ is the higher degree of $Z_x$. The resulting equation reads,
\begin{equation}\label{eq:GdZx}
  \gamma_0^2 \left(\gamma_0^2 (x^2-Z_z^2)^2-16 \beta^2-4 (x^2+Z_z^2)/\gamma_0\right)-2 \gamma_0 \Omega_B^2 \left(2 \beta^2+\gamma_0 (x^2+Z_z^2)-2\right)+\Omega_B^4 = 0.
\end{equation}
This equation can be solved exactly in $x$. The solution yielding growing modes reads,
\begin{equation}\label{Eq:xuhl}
  \delta^2=\frac{\gamma_0^2 \left(\Omega_B^2+\gamma_0 \left(-2 \beta ^2+\gamma_0 Z_z^2+2\right)\right)-2 \sqrt{\gamma_0^6 \left(\beta ^4+\beta ^2 \left(2-2 \gamma_0 Z_z^2\right)+Z_z^2 \left(2 \gamma_0+\Omega_B^2\right)+1\right)}}{\gamma_0^4},
\end{equation}
where the mode grows when $\delta^2<0$. The equation $\delta(Z_z)=0$ can also be solved, and the growth-rate $\delta$  is found positive only for,
\begin{equation}
 Z_{z1} \equiv \frac{\sqrt{\Omega_B^2-4 \beta ^2 \gamma_0}}{\gamma_0} < Z_z <
 \frac{\sqrt{\Omega_B^2 + 4 \gamma_0}}{\gamma_0} \equiv  Z_{z2}.
 \end{equation}
While the Weibel instability has not vanished, i.e., Eq. (\ref{Eq:Weibel0}) is not satisfied, $Z_{z1}$ is purely imaginary so that this growth-rate is finite between $Z_z=0$ and $Z_{z2}$. For $\Omega_B$ larger than $2\beta \sqrt{\gamma_0}$,  Weibel vanishes, $Z_{z1}>0$,  and we obtain the upper-hybrid-like instability. In this regime,  the growth-rate reaches a maximum for,
\begin{equation}
Z_{z,max} \sim \frac{\Omega_B}{\gamma_0}.
\end{equation}
By replacing $Z_z$ by this value in Eq. (\ref{Eq:xuhl}), one finds for the maximum growth-rate of these upper-hybrid-like modes,
\begin{equation}\label{Eq:uhl}
  \delta_U^2=2\frac{1 + \gamma_0 \Omega_B^2 - \sqrt{4 \gamma_0^4+\gamma_0^2 \left(\Omega_B^4-4\right)+2 \gamma_0 \Omega_B^2+1}}{\gamma_0^3},
\end{equation}
with the following limit expressions,
\begin{eqnarray}\label{Eq:uhllimit}
 \lim_{\gamma_0 \rightarrow \infty} \delta_U &=& \frac{2}{\sqrt{\gamma_0}} \nonumber \\
 \lim_{\Omega_B \rightarrow \infty} \delta_U &=& \frac{2\beta}{\Omega_B}\nonumber
\end{eqnarray}

The growth-rate at large $\Omega_B$ is therefore independent of the beams energy $\gamma_0$. The same conclusion has already been reached for the case of two counter-streaming electron beams with guiding magnetic field \cite{Godfrey1975,BretPoPMagne}.

\subsection{Region 3: Two-stream and transverse instability Regions}\label{sec:OTS}
With $Z_z=Z_\perp=0$, the tensor $\mathcal{T}$ turns diagonal and the tensor elements (\ref{eq:Telt}) read,
\begin{eqnarray}
  \mathcal{T}_{11}&=& 1-\frac{Z_z^2}{\beta^2 x^2}+ \frac{1}{x^2} \frac{4 \gamma_0 \Omega_B^2 \left(x^2+Z_z^2\right)-4 \gamma_0^3 \left(x^2-Z_z^2\right)^2}{\gamma_0^4 \left(x^2-Z_z^2\right)^2-2 \gamma_0^2 \Omega_B^2 \left(x^2+Z_z^2\right)+\Omega_B^4},  \nonumber \\
 \mathcal{T}_{22}&=\mathcal{T}_{11} , \nonumber \\
  \mathcal{T}_{33}&=& 1-\frac{4 \left(x^2+Z_z^2\right)}{\gamma_0^3 \left(x^2-Z_z^2\right)^2} .\nonumber
\end{eqnarray}
The equation $\mathcal{T}_{33}=0$ pertains to the two-stream instability. It can be analyzed exactly. The maximum growth-rate is,
\begin{equation}\label{eq:TS}
  \delta_{TS} = \frac{1}{\sqrt{2}\gamma_0^{3/2}},~~\mathrm{for}~~Z_z= \frac{\sqrt{3/2}}{\gamma_0^{3/2}}.
\end{equation}
This result is rigourously independent of $\Omega_B$ because the two-stream instability has the particles oscillating along the flow. If the magnetic field is flow-aligned, as is the case here, the Lorentz force vanishes and the instability remains unchanged by the field.

The equation $\mathcal{T}_{11}=0$ pertains to the transverse instability already investigated in Refs. \cite{TautzB02005,Lazar2008}. By Taylor-expanding the dispersion equation up to second order in $x$, one finds the instability occurs for,
\begin{equation}\label{eq:ZzOmode}
  \frac{\Omega_B}{\gamma_0} \sqrt{1-\frac{4 \beta^2 \gamma_0}{\Omega_B}} \lesssim Z_z \lesssim \frac{\Omega_B}{\gamma_0},
\end{equation}
with a maximum growth-rate reached for $Z_{z,max}$ towards the middle of the interval. From the second order equation derived from the previous Taylor expansion, and the value of $Z_{z,max}$, one can derive the following expression for the maximum growth-rate (we set $\beta= 1$),
\begin{equation}\label{eq:Omode}
  \delta_T^2 = \frac{\Omega_B (5 \varphi+3 \Omega_B)-6 \gamma_0}
  {2 \gamma_0 \Omega_B (\varphi+2 \Omega_B)+\Omega_B^3 (3 \varphi+5 \Omega_B)-10 \gamma_0^2}, ~~\mathrm{with}~~\varphi = \sqrt{\Omega_B^2-4 \beta^2 \gamma_0}.
  \end{equation}
For large $\Omega_B$, we get the following expansion,
\begin{equation}\label{eq:Omode-app}
\delta_T \sim \frac{1}{\Omega_B}.
\end{equation}

We see from Fig. \ref{fig:hie}(B) that the fastest growing mode has $Z_z=Z_\perp=0$ in Region 3. Yet, this is not enough to pinpoint the kind of mode we are dealing with, since there are two such modes.

\begin{figure}
  \includegraphics[width=0.4\textwidth]{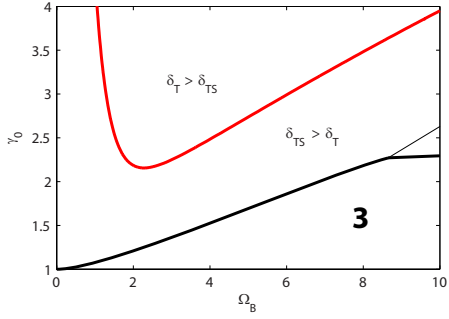}
  \caption{(Color online) Numerical comparison of the maximum growth-rates of the two-stream and the transverse instabilities. Below the red line, the two-stream instability grows faster than the transverse instability. Region 3 is therefore governed by the two-stream instability. }\label{fig:OvsTS}
\end{figure}

In order to spot the fastest growing mode in this region, we need therefore to compare   $\delta_{TS}$ and $ \delta_T$. While Eq. (\ref{eq:TS}) is an exact expression for $\delta_{TS}$, Eq. (\ref{eq:Omode-app}) is only valid at large $\Omega_B$. A numerical comparison of the two growth-rates, valid therefore at any $\Omega_B$, is displayed on Figure \ref{fig:OvsTS} together with the extent of Region 3. Clearly, the two-stream instability is the fastest growing one in this region.

\begin{figure}
  \includegraphics[width=\textwidth]{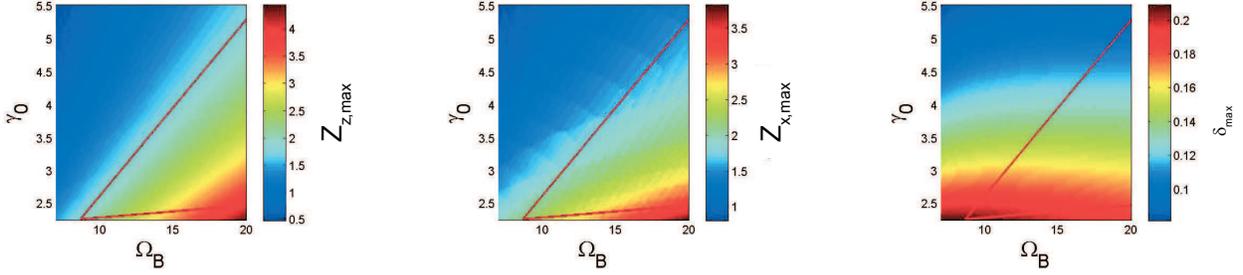}
  \caption{(Color online) Numerical evaluation of the parallel and perpendicular components of the oblique mode, together with its growth-rate, in terms of $\Omega_B$ and $\gamma_0$. This mode governs the full unstable spectrum between to two red lines.}\label{fig:ob}
\end{figure}

\subsection{Region 4: Oblique Region}
The unstable spectrum is here dominated by oblique-like modes such as the one pictured on Fig. \ref{fig:maps}(4). Our study is here mainly numerical. Noting that the most unstable mode has $(Z_z,Z_x) \sim (\Omega_B/2\gamma_0,\Omega_B/2\gamma_0)$, one can conduct an accurate numerical search of its characteristics.

The result is pictured on Fig. \ref{fig:ob}, where the red lines enclose the domain where this mode govern the full spectrum. Within, this domain, the growth-rate is reasonably well approximated by,
\begin{equation}\label{eq:ob}
  \delta_O \sim \frac{1}{2\gamma_0}.
\end{equation}

\section{Frontiers equations}\label{sec4}
Having expressed the largest growth-rate in each region, we can now determine the frontiers between these domains.

\subsection{Weibel/UHL frontier (1/2)}
We here look for the equation of the frontier between Region 1, governed by the Weibel instability, and Region
 2, governed by upper-hybrid-like modes. Such a determination comes down to comparing Eqs. (\ref{Eq:Weibel}) and (\ref{Eq:uhl}). The comparison of the two growth-rates can be worked out exactly. They are equal for,
\begin{equation}\label{eq:front12}
  \gamma_0=\frac{3}{16} \left(\Omega_B^2+\sqrt{\Omega_B^4+128/3}\right),
\end{equation}
with the following limits,
\begin{eqnarray}
\gamma_0 &=& \sqrt{\frac{3}{2}} + \frac{3}{16}\Omega_B^2,~~\mathrm{for}~~\Omega_B \ll 1, \nonumber \\
\gamma_0 &=& \frac{3}{8}\Omega_B^2,~~\mathrm{for}~~\Omega_B \gg (128/3)^{1/4} \sim 2.5
\end{eqnarray}

\subsection{UHL/two-stream frontier (2/3)}
We now turn to the equation of the frontier between regions 2 and 3, comparing Eqs. (\ref{Eq:uhl}) and (\ref{eq:TS}). Here, the equation can be solved exactly in terms of $\Omega_B$, giving equality of the two growth-rates for,
\begin{equation}\label{eq:23Grand}
  \Omega_B=\frac{4\sqrt{\gamma_0^4-\gamma_0^2-9/64}}{ \sqrt{2\gamma_0}}.
\end{equation}
For $\gamma_0 \gg 1$, this gives,
\begin{equation}\label{eq:front23}
  \gamma_0=\frac{\Omega_B^{2/3}}{2}.
\end{equation}
Equation (\ref{eq:23Grand}) cannot be extended down to $\Omega_B=0$ because the analytical expression (\ref{Eq:uhl}) for the upper-hybrid-like growth-rate is only valid for $\Omega_B > 2 \beta\sqrt{\gamma_0}$. For smaller values of $\Omega_B$, one can come back to the UHL/Weibel growth-rate expression (\ref{Eq:xuhl}) and directly search for its maximum in terms of $Z_z$. One finds the maximum growth-rate is exactly reached for,
\begin{equation}\label{eq:ZzMax2}
  Z_z^2= \left(3-\beta ^2\right) \gamma_0+\Omega_B^2 \frac{\left(1-3 \beta ^2\right) \gamma_0+\Omega_B^2}
  {\gamma_0^2 (\Omega_B^2 + 2/\gamma_0)}.
\end{equation}
Replacing then $Z_z$ by this expression in Eq. (\ref{Eq:uhl}), and Taylor expanding the result near $(\Omega_B,\gamma_0)=(0,1)$, one finds in this regime,
\begin{equation}
 \delta_U-\delta_{TS}= \frac{(5-7 \gamma_0) \Omega_B^2}{8 \sqrt{2}} + 2 \sqrt{2} (\gamma_0-1) .
\end{equation}
For $\Omega_B \ll 1$, the growth-rate $\delta_U$ is eventually found larger than $\delta_{TS}$ for,
\begin{equation}
 \gamma_0 > 1 + \frac{\Omega_B^2}{16}.
\end{equation}

\subsection{UHL/Oblique and Two-stream/Oblique frontiers (2/4 and 3/4)}
Owing to the weak analytical expression (\ref{eq:ob}) available for the oblique modes, an accurate description of these frontiers has to be numerical. It is nevertheless interesting to check the frontiers set by our analytical expression (\ref{eq:ob}).

Regarding the UHL/Oblique frontier, comparing Eqs. (\ref{Eq:uhl}) and (\ref{eq:ob}) gives,
\begin{equation}\label{eq:front24}
  \Omega_B=\frac{\sqrt{256 \gamma_0^3-257 \gamma_0-16}}{4 \sqrt{\gamma_0}} \sim 4\gamma_0~~\mathrm{for}~~\gamma_0 \gg 1.
\end{equation}

Turning now to the Oblique/Two-stream frontier, comparing Eqs. (\ref{eq:TS}) and (\ref{eq:ob}) simply gives,
\begin{equation}\label{eq:front34}
  \gamma_0 = 2.
\end{equation}

\begin{figure}
  \includegraphics[width=\textwidth]{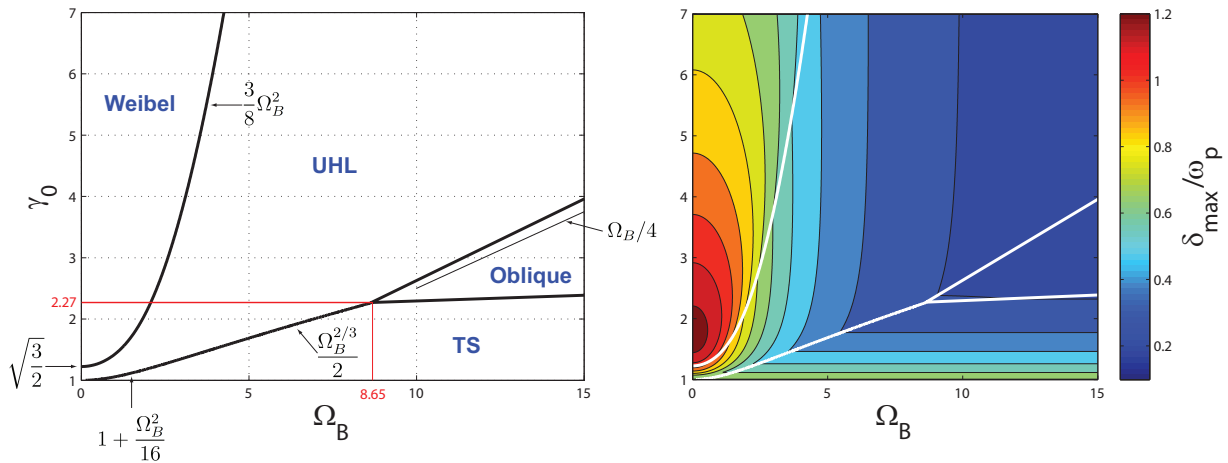}
  \caption{(Color online) Summary of the main results gathered in this work}\label{fig:recap}
\end{figure}

As evidenced by the numerically determined hierarchy map on Figs. \ref{fig:maps} \& \ref{fig:recap}, these expressions above are only qualitatively correct.

\begin{figure}
  \includegraphics[width=\textwidth]{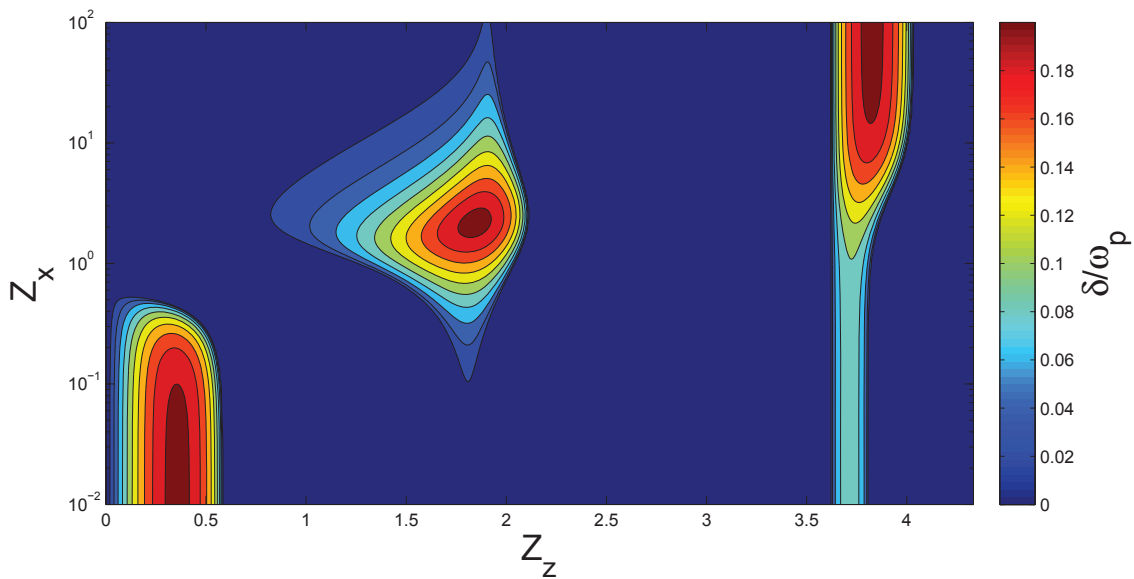}
  \caption{(Color online) Growth rate in terms of the wave-vector at the ``triple point'', $\Omega_B=8.65$ and $\gamma_0=2.27$. The two-stream, the oblique and the upper-hybrid-like instabilities grow exactly the same rate.}\label{fig:triple}
\end{figure}

\subsection{Triple point}
Our hierarchy analysis unravels the existence of a ``triple point'' in the $(\Omega_B,\gamma_0)$ parameter space, where the two-stream, the oblique and the upper-hybrid-like instabilities grow exactly the same rate. This point of the parameters phase space if found for,
\begin{eqnarray}\label{eq:triple}
  \Omega_B&=&8.65, \nonumber \\
  \gamma_0&=&2.27.
\end{eqnarray}
Such a situation is pictured on Fig. \ref{fig:triple} where a logarithmic scale as been applied to the $Z_x$ axis only. We find here the 3 different modes grow at about $0.2\omega_p^{-1}$.

\section{Conclusion}
The hierarchy of the instabilities involved in the interaction of two counter-streaming pair beams in a guiding magnetic field has been worked out. The two beams are considered cold and symmetric, so that only two parameters control the system: the initial Lorenz factor of the beams, and the strength of the external magnetic field.

The result evidences 4 different kinds of instabilities governing 4 different regions of the $(\Omega_B, \gamma_0)$ space. Figure \ref{fig:recap} summarizes most of the results gathered in this work. For $\Omega_B=0$, we find upper-hybrid-like modes govern the spectrum provided $\gamma_0<\sqrt{3/2}$, while the Weibel instability dominates otherwise. This correctly merges with the result obtained previously for the case of two symmetric, unmagnetized counter-streaming electron beams \cite{BretPoP2013} (note that the upper-hybrid-like modes were labeled ``oblique'' in Ref. \cite{BretPoP2013}).

Figure \ref{fig:recap} also shows that the largest growth rate that can be obtained for any couple $(\Omega_B,\gamma_0)$ is found for $\Omega_B=0$, in the Weibel region. In this case, Eq. (\ref{Eq:Weibel}) gives $\delta_W=2\beta/\sqrt{\gamma_0}$, which reaches the extremum $\delta_{W, max}=8/3^{3/2}\sim 1.54$, for $\gamma_0=\sqrt{3}$ (or $\beta=\sqrt{2/3}$).

The cold regime presently studied allows to reach a significant number of analytical results. In the case of counter-streaming electron beams, the cold regime turned out to be a reliable guide for the temperature dependant study \cite{BretPoPHierarchie,BretPRL2008,BretPRE2010}. There is little doubt that here, the present study can also be a guide for future temperature dependant investigations. The kinetic version of our hierarchy map may however prove far more difficult to work out than its un-magnetized counterpart, precisely due to the presence of a magnetic field. Indeed, the calculation of the growth-rate over the full \textbf{k}-space has only been performed so far for an electron beam passing through a Maxwellian, non-relativistic plasma \cite{timofeev2009,Timofeev2013}. Progresses towards a full relativistic and kinetic treatment, like the one achieved in Ref. \cite{BretPRE2010} for the un-magnetized case, would be desirable before the present problem can be tackled kinetically.

\section{Acknowledgments}
This work was supported by grant ENE2013-45661-C2-1-P from the Ministerio de Educaci\'{o}n y Ciencia, Spain
and grant PEII-2014-008-P from the Junta de Comunidades de Castilla-La Mancha.

\appendix
\section{Tensor elements}\label{app}
The tensor $\mathcal{T}$ defined by Eq. (\ref{eq:Max}) reads in terms of the dimensionless variables (\ref{eq:var}),
\begin{equation}
  \mathcal{T} =
  \begin{bmatrix}
     \mathcal{T}_{11} & 0  & \mathcal{T}_{13}\\
     0 & \mathcal{T}_{22}  & 0\\
     \mathcal{T}_{31} & 0  & \mathcal{T}_{33}
  \end{bmatrix},
\end{equation}
where,
\begin{eqnarray}\label{eq:Telt}
\mathcal{T}_{11} &=& 1-\frac{Z_z^2}{\beta^2 x^2}+\frac{4\gamma_0}{x^2} \frac{\Omega_B^2 \left(x^2+Z_z^2\right)-\gamma_0^2 \left(x^2-Z_z^2\right)^2}
{\Omega_B^4-2 \gamma_0^2 \Omega_B^2 \left(x^2+Z_z^2\right)+\gamma_0^4 \left(x^2-Z_z^2\right)^2} \\
\mathcal{T}_{22} &=& 1-\frac{Z_z^2+Z_x^2}{\beta^2 x^2} + \frac{4\gamma_0}{x^2}\frac{ x^2 \left(\Omega_B^2+2 \gamma_0^2 Z_z^2\right)+\Omega_B^2 Z_z^2-\gamma_0^2 x^4-\gamma_0^2 Z_z^4}
{\gamma_0^4 x^4-2 \gamma_0^2 x^2 \left(\Omega_B^2+\gamma_0^2 Z_z^2\right)+\left(\Omega_B^2-\gamma_0^2 Z_z^2\right)^2}
  \nonumber \\
\mathcal{T}_{33} &=&  1+ \frac{1}{x^2}\left[ 4 \gamma_0 Z_x^2 \left(\frac{ \Omega_B^2-\gamma_0^2 \left(x^2+Z_z^2\right)}{\Omega_B^4-2 \gamma_0^2 \Omega_B^2 \left(x^2+Z_z^2\right)+\gamma_0^4 \left(x^2-Z_z^2\right)^2}-\frac{1}{4 \gamma_0\beta^2}\right) -\frac{4 x^2 \left(x^2+Z_z^2\right)}{\gamma_0^3 \left(x^2-Z_z^2\right)^2}
\right]
\nonumber \\
\mathcal{T}_{13} &=& \frac{Z_x Z_z}{x^2} \left[
 \frac{1}{\beta^2}-4 \gamma_0\frac{ \Omega_B^2+\gamma_0^2 \left(x^2-Z_z^2\right)}{\Omega_B^4-2 \gamma_0^2 \Omega_B^2 \left(x^2+Z_z^2\right)+\gamma_0^4 \left(x^2-Z_z^2\right)^2}
 \right] \nonumber \\
 \mathcal{T}_{13} &=& \mathcal{T}_{31}  \nonumber
\end{eqnarray}


\end{document}